\documentclass[prl,aps,twocolumn,nopacs,superscriptaddress,nofootinbib]{revtex4-2}

\usepackage{amsmath}
\usepackage{amssymb}
\usepackage{amsfonts}
\usepackage{bm} 
\usepackage{bbm}
\usepackage{braket}
\usepackage{color}
\usepackage{comment}
\usepackage{dcolumn} 
\usepackage{dsfont}
\usepackage{enumerate}
\usepackage{epsfig}
\usepackage{esint}
\usepackage[T1]{fontenc}
\usepackage{framed}
\usepackage{gensymb}
\usepackage{graphicx} 
\usepackage[colorlinks,linkcolor=blue,citecolor=blue,urlcolor=blue,hyperindex,driverfallback=dvipdfm]{hyperref}
\usepackage{indentfirst}
\usepackage{lmodern}
\usepackage{mathrsfs}
\usepackage{mathtools}
\usepackage{multirow}
\usepackage{psfrag}
\usepackage{pst-all}
\usepackage{soul}
\usepackage{units}
\usepackage{xcolor}
\usepackage{xspace}

\newcommand{\abs}[1] {\mathopen{}\left|#1\right|\mathclose{}}

\newcommand{\ccpar}[1] {\mathopen{}\left(#1\right)\mathclose{}}
\newcommand{\sqpar}[1] {\mathopen{}\left[#1\right]\mathclose{}}
\newcommand{\clpar}[1] {\mathopen{}\left\{#1\right\}\mathclose{}}
\newcommand{\av}[1] {\left\langle #1 \right\rangle}
\def\ii{{\rm i}}  \def\ee{{\rm e}}
  
\def\Ree{{\rm Re}}  \def\Imm{{\rm Im}}

\def\rb{{\bf r}}

\def\Eb{{\bf E}}          
    
\def\db{{\bf d}}    
\def\rp{r_{\rm p}}    
    \def\EF{{E_{\rm F}}}  
\def\eps{\epsilon}  \def\vep{\varepsilon}
\def\ww{\omega}  
\def\wL{\omega_{\rm L}}
\def\epsa{\epsilon_{\rm a}}  \def\epsb{\epsilon_{\rm b}}
\def\rhot{\tilde{\rho}}

    \def\Gm{{\mathcal{G}}}
    
  \def\Vm{{\mathcal{V}}}

\begin{document}

\title{Optoelectronic control of atomic bistability with graphene}

\author{Mikkel~Have~Eriksen}
\affiliation{Center for Nano Optics, University of Southern Denmark, Campusvej 55, DK-5230 Odense M, Denmark}

\author{Jakob~E.~Olsen}
\affiliation{Faculty of Engineering, University of Southern Denmark, Campusvej 55, DK-5230 Odense M, Denmark}

\author{Christian~Wolff}
\affiliation{Center for Nano Optics, University of Southern Denmark, Campusvej 55, DK-5230 Odense M, Denmark}

\author{Joel~D.~Cox}
\email[Joel~D.~Cox: ]{cox@mci.sdu.dk}
\affiliation{Center for Nano Optics, University of Southern Denmark, Campusvej 55, DK-5230 Odense M, Denmark}
\affiliation{Danish Institute for Advanced Study, University of Southern Denmark, Campusvej 55, DK-5230 Odense M, Denmark}

\begin{abstract}
We explore the emergence and active control of optical bistability in a two-level atom near a graphene sheet. Our theory incorporates self-interaction of the optically-driven atom and its coupling to electromagnetic vacuum modes, both of which are sensitive to the electrically-tunable interband transition threshold in graphene. We show that electro-optical bistability and hysteresis can manifest in the intensity, spectrum, and quantum statistics of the light emitted by the atom, which undergoes critical slow-down to steady-state. The optically-driven atom-graphene interaction constitutes a platform for active control of driven atomic systems in quantum coherent control and atomic physics.
\end{abstract}

\date{\today}
\maketitle




Coherent optical control of atomic systems enables fundamental explorations of quantum physics while promising disruptive applications in diverse fields, ranging from information and communication technologies to optical sensing and metrology \cite {streltsov2017colloquium}. In this context, nanophotonic architectures that enhance atom-photon interactions offer a robust and scalable platform upon which to develop next-generation integrated photonic devices \cite{lodahl2015interfacing,chang2018colloquium}. Metal nanostructures supporting plasmons---collective excitations in the free electron plasma---have been widely explored as subwavelength light-focusing elements in hybrid systems, where the combined broad spectral response of a plasmonic resonator and the narrow linewidth of a few-level quantum light emitter (e.g., a quantum dot) are predicted to enable phenomena such as nonlinear Fano effects \cite{zhang2006semiconductor,ridolfo2010quantum}, 
optical bistability \cite{artuso2008optical,artuso2010strongly,carreno2018nonlinear}, optical hysteresis \cite{malyshev2011optical,malyshev2012condition,arrieta2014modelling}, excitonic population transfer \cite{anton2012plasmonic}, and enhanced resonance fluorescence \cite{carreno2013resonance,mohammadzadeh2019resonance}; a salient feature in these and related studies are nonlinear dynamics emerging from the self-interaction of the atomic transition dipole mediated by plasmon resonances.

While plasmons in noble metals provide nanoscopic light focusing, they suffer from large intrinsic Ohmic loss and cannot easily be tuned in an active manner. These limitations are partly alleviated in highly-doped graphene, which hosts long-lived and actively-tunable plasmon resonances that strongly concentrate light \cite{gonccalves2016introduction}, thus presenting new opportunities to control atom-light interactions on the nanoscale \cite{koppens2011graphene,manjavacas2012temporal,forati2014graphene,chang2017constructing}. Unfortunately, graphene plasmons are limited by charge carrier doping, with achievable Fermi levels $\EF\lesssim 1$\,eV restricting resonances to the terahertz and infrared spectral range lying well-below the operational frequencies of robust quantum light sources \cite{garcia2014graphene,cox2018nonlinear}. 
Nevertheless, the carbon monolayer exhibits an impressive light-matter interaction associated with optical excitation of electrons between conical valence and conduction bands, giving rise to a broadband 2.3\% light absorption at energies beyond the electrically-tunable $2\EF$ threshold \cite{wang2008gate,gonccalves2016introduction}. In the context of coherent optical control, recent experiments confirm that optoelectronic tunability of the graphene interband response can be harnessed to manipulate quantum light emission \cite{lee2014switching,tielrooij2015electrical}, also enabling fast dynamical control of strong near-field interactions that produce $\gtrsim1000$-fold enhancement in the decay rate of erbium emitters \cite{cano2020fast}.

Considerable efforts have been made to control spontaneous emission dynamics of excited quantum light emitters using the electrical tunability of graphene, yet far fewer investigations have explored the consequences of interfacing optically-driven atomic systems with an actively-tunable nanophotonic environment \cite{forati2014graphene,arrieta2014modelling}. In particular, changes in the local photonic density of states experienced by an atom in the spectral neighborhood of its transition frequency can impact both the optically-induced self-interaction and mesoscopic quantum electrodynamic phenomena that manifest from vacuum fluctuations \cite{scully1999quantum}, such as the Purcell effect and Lamb shift \cite{koppens2011graphene,chang2017constructing}. The dynamics of a driven two-level atom experiencing all the aforementioned phenomena is hitherto unexplored, even in studies of atom-plasmon interactions. 

In this letter, we theoretically explore the nonlinear response of an optically-driven two-level atom near an electrically-tunable graphene sheet that mediates optical and vacuum-induced light-matter interactions. We focus on optical bistability that emerges from the feedback of the optically-induced atomic dipole produced by the carbon monolayer, which we demonstrate can be harnessed to actively switch the hybrid system into different metastable states. The interband transition threshold in the graphene sheet simultaneously impacts the spontaneous emission rate (Purcell effect) and atomic transition frequency (Lamb shift), leading to complex dynamics in the response of the two-level system. Our findings motivate further studies of quantum electrodynamic effects in atomic bistability, while offering a prescription for active and \textit{in situ} modification of quantum optical states in optical lattices and integrated nanophotonic platforms.

\begin{figure*}
    \centering
    \includegraphics[width=\textwidth]{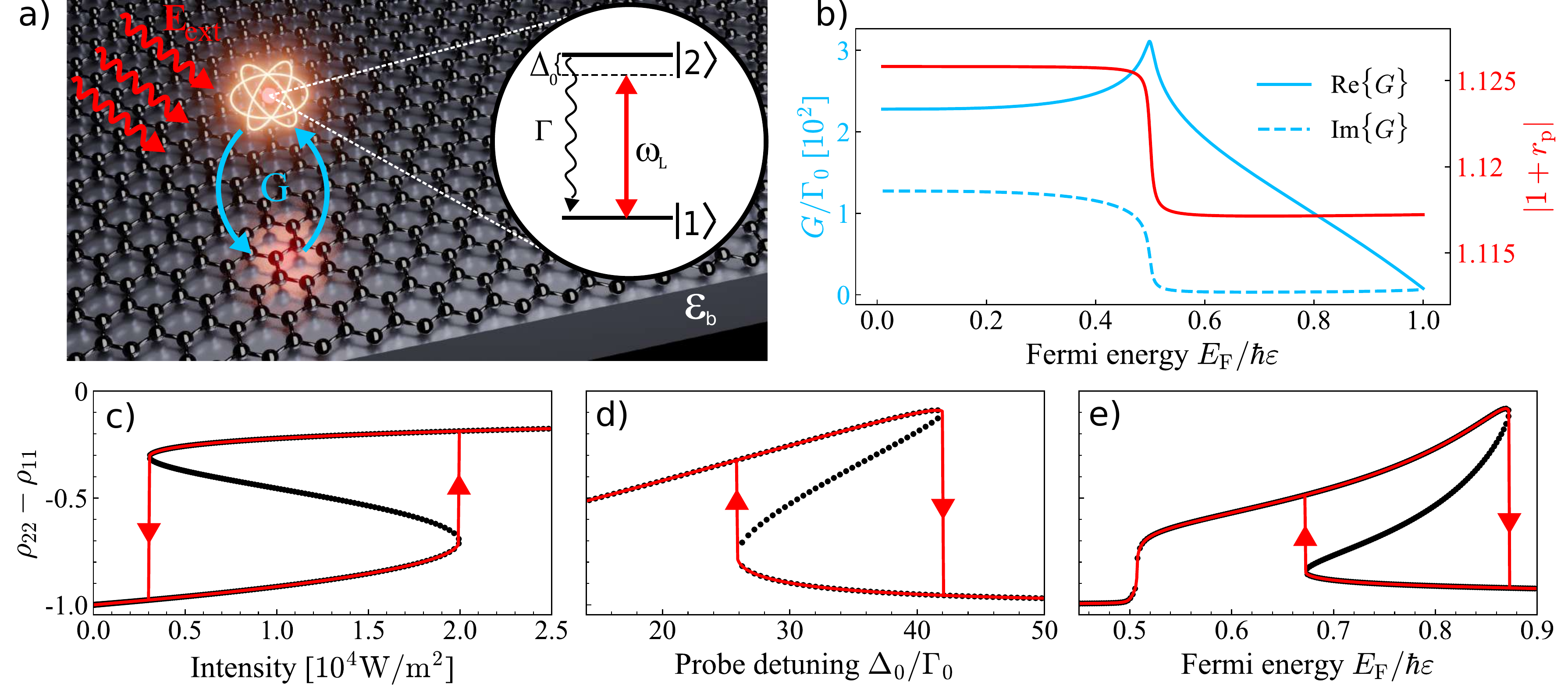}
    \caption{\textbf{Optical bistability in an atom interfacing graphene.} (a) Schematic illustration of a two-level atom (TLA) with ground state $\ket{1}$ and excited state $\ket{2}$, placed at a distance $z$ above an extended graphene sheet encapsulated in media with dielectric permittivity $\epsa$ above and $\epsb$ below. (b) Self-interaction parameter $G$ (left vertical axis) governing the dynamics of a TLA with transition energy $\hbar\vep=1.0$\,eV and vacuum decay rate $\Gamma_0$ placed $z=12$\,nm above a graphene sheet as the Fermi energy $\EF$ is varied; the reflection coefficient $r_{\rm p}$ for p-polarized light (right axis) enters both $G$ and the effective Rabi frequency $\Omega$. (c-e) The steady-state TLA population difference $Z=\rho_{22}-\rho_{11}$ is plotted in red curves corresponding to time-domain simulations obtained by adiabatically sweeping (c) the external field intensity $I^{\rm ext}$ at a distance $z=17$\,nm, (d) the detuning $\Delta_0=\wL-\vep$ at $z=17$\,nm, and (e) the Fermi energy $\EF$ at $z=12$\,nm, and exhibit hysteresis indicated by the arrows; black dots correspond to solutions of Eq.\ \eqref{eq:Z_SS}. Unless explicitly varied, the results presented in (b-e) correspond to parameters $\epsa=1$, $\epsb=1.6$, $I^{\rm ext}=10^4$\,W/m$^2$, $\hbar\Delta_0=8$\,$\mu$eV, and $\EF=0.51$\,eV, while the TLA decay rate and transition dipole moment are $\Gamma_0 \approx 0.38$\,ns$^{-1}$ and $d = 1$\,$e\cdot$nm, respectively, and the broadening associated with inelastic scattering in graphene is $\hbar\tau^{-1} = 0.01$\,eV.} %
    \label{fig1}
\end{figure*}

We consider a generic two-level atomic system (e.g., an atom or quantum dot) positioned at $\rb=(x,y,z)$ above a graphene sheet extended in the $z=0$ plane and interfacing homogeneous dielectric media with permittivity $\eps_{\rm a}$ above and $\eps_{\rm b}$ below, respectively, as depicted schematically in Fig.\ \ref{fig1}a. The Hamiltonian of the two-level atom (TLA) is expressed as
\begin{align}
    \mathcal{H} &= \hbar\sum_{j=1}^2\vep_j\ket{j}\bra{j} + \hbar \int_0^\infty d\ww \ww \int d^3\rb \hat{\bf f}_\ww^\dagger(\rb)\cdot\hat{\bf f}_\ww(\rb)  \nonumber \\
    &-\hat{\db}\cdot\sqpar{\Eb(\rb,\wL)\ee^{-\ii\wL t} + \int_0^\infty d\ww \hat{\Eb}_{\rm R}(\rb,\ww) + {\rm h.c.}} ,  \label{eq:H}
\end{align}
where the first term is the Hamiltonian of the bare atom comprised of states $\ket{j}$ with energies $\hbar\vep_j$ for $j\in\{1,2\}$, the second term corresponds to the vacuum radiation field, expressed in terms of bosonic field operators $\hat{\bf f}_\ww^\dagger$ ($\hat{\bf f}_\ww$) that create (annihilate) photons, and the final term describes the atom coupling with the classical monochromatic field $\Eb^{\rm ext}\ee^{-\ii\wL t}+\text{c.c.}$ and quantized radiation field operator $\hat{\Eb}_{\rm R}$ through their projection on the dipole operator $\hat{\db}$ \cite{carreno2013resonance, forati2014graphene, antao2021two}.

The classical field is comprised of the external field, its reflection by the graphene sheet, and the image field produced by the TLA dipole in graphene, so that
\begin{equation} \label{eq:E}
    \Eb(\rb,\wL) = \tilde{\epsilon}^{-1}\clpar{\sqpar{1 + \rp(0,\wL)}\Eb^{\rm ext} + \wL^2 \mu_0 \Gm_{\omega_{\rm L}}(\rb, \rb) \cdot \langle\hat{\db}\rangle} ,
\end{equation}
where
\begin{equation}
    \rp(Q,\ww) = \frac{\epsb k_{{\rm b},z} - \epsa k_{{\rm a},z} + k_{{\rm a},z}k_{{\rm b},z}\sigma/(\ww\eps_0)}{\epsb k_{{\rm a},z} + \epsa k_{{\rm b},z} + k_{{\rm a},z}k_{{\rm b},z}\sigma/(\ww\eps_0)}
\end{equation}
is the reflection coefficient for p-polarized light, expressed in terms of the normal wave vector components $k_{m,z}=\sqrt{\eps_{m}\ww^2/c^2-Q^2}$ in medium $m\in\{{\rm a},{\rm b}\}$, the conserved in-plane wave vector $Q$, and the graphene surface conductivity $\sigma$; these quantities also enter the reflected part of the Green's tensor $\Gm_\ww(\rb,\rb')$ that mediates the self-interaction of the dipole $\langle\hat{\db}\rangle$, where $\av{\dots}$ denotes the quantum mechanical average. Note that Eq.\ \eqref{eq:E} assumes an external field impinging normally to the graphene sheet, while the prefactor $\tilde{\epsilon}^{-1}$ accounts for possible dielectric screening by the internal structure of the atom relative to its host environment \cite{malyshev2011optical}.

The quantum field operator of the inhomogeneous photonic environment can be expressed in terms of the classical Green's function as \cite{scheel2008macroscopic,forati2014graphene}
\begin{equation}
    \hat{\Eb}_{\rm R}(\rb,\ww) = \ii\sqrt{\frac{\hbar}{\pi\eps_0}}\frac{\ww^2}{c^2}\int d^3\rb' \sqrt{\Imm\clpar{\chi_\ww(\rb')}}\Gm_\ww(\rb,\rb')\cdot\hat{\bf f}_\ww(\rb') ,
\end{equation}
where $\chi$ is the susceptibility of the dielectric background. Following the procedure described in the Supplementary Information (SI), we trace over the photonic reservoir to form a master equation for the density matrix $\hat{\rho}$ that governs the TLA dynamics in the interaction picture:
\begin{align}
    \dot{\hat{\rho}} = &-\frac{\ii}{\hbar}\sqpar{\Vm+\hbar\delta\ww\ket{2}\bra{2},\hat{\rho}}  \nonumber \\
    &+\frac{\Gamma}{2}\ccpar{2\ket{1}\bra{2}\hat{\rho}\ket{2}\bra{1} - \ket{2}\bra{2}\hat{\rho} - \hat{\rho}\ket{2}\bra{2}} ,  \label{eq:rho_eom}
\end{align}
where
\begin{equation} \label{eq:V_AL}
    \Vm = -\sqpar{\db\cdot\Eb(\rb,\wL)\ee^{-\ii\wL t}+{\rm c.c.}}\ccpar{\ket{1}\bra{2}\ee^{-\ii\vep t}+{\rm h.c.}}
\end{equation}
is the atom-light interaction Hamiltonian,
\begin{equation} \label{eq:Purcell}
    \Gamma = \Gamma_0 + \frac{2\mu_0}{\hbar} \vep^2 \Imm\clpar{\db \cdot \Gm_{\vep}(\rb, \rb) \cdot \db}
\end{equation}
is the spontaneous emission rate, with $\Gamma_0=\vep^3\abs{\db}^2/3\pi\eps_0\hbar c^3$ denoting the vacuum emission rate \cite{novotny2012principles}, and
\begin{equation} \label{eq:Lamb}
    \delta \omega = \frac{ \mu_0}{\pi \hbar}\mathcal{P} \int_0^{\infty}d\omega \omega^2  \frac{\db \cdot \text{Im}\{\Gm_\ww(\textbf{r}_0, \textbf{r}_0) \} \cdot \db}{\vep - \omega}
\end{equation}
quantifies the Lamb shift as a Cauchy principal value integral. In Eqs.\ \eqref{eq:V_AL}-\eqref{eq:Lamb}, we have introduced the atomic transition frequency $\vep\equiv\vep_2-\vep_1$ and transition dipole moment $\db\equiv\braket{1|\db|2}=\braket{2|\db|1}$. Inserting the Hamiltonian of Eq.\ \eqref{eq:H} into Eq.\ \eqref{eq:rho_eom} and writing $\langle\hat{\db}\rangle=\text{Tr}\{\hat{\db}\hat{\rho}\}=\db\rho_{21}+{\rm c.c.}$, we obtain the familiar equations of motion for the density matrix elements $\rho_{jj'}=\braket{j|\hat{\rho}|j'}$ in the rotating-wave approximation
\begin{subequations} \label{eq:rho_ij_eom}
\begin{align}
    \frac{\partial\rho_{11}}{\partial t} &= \Gamma\rho_{22}+\ii\ccpar{\Omega^* + G^*\rhot_{12}}\rhot_{21}-\ii\ccpar{\Omega + G\rhot_{21}}\rhot_{12}  \\
    \frac{\partial\rhot_{21}}{\partial t} &= \ccpar{\ii \Delta - \gamma}\rhot_{21}-\ii\ccpar{\Omega + G\rhot_{21}}\ccpar{\rho_{22}-\rho_{11}},
\end{align}
\end{subequations}
where $\Delta\equiv \wL - \vep + \delta\omega$ is the effective detuning parameter (including the Lamb shift), $\rhot_{12}=(\rhot_{21})^*=\rho_{12}\ee^{-\ii\ww t}$ are the coherence elements transformed to a frame oscillating with the external field,
$\Omega=(1 + r_{\rm p})\db\cdot{\Eb}^{\rm ext}/(\tilde{\eps}\hbar)$ is the Rabi frequency, renormalized from its free space value $\db\cdot\Eb^{\rm ext}/\hbar$, and
$G = \ww_{\rm L}^2\mu_0\db\cdot\Gm_{\omega_{\rm L}}(\rb,\rb)\cdot\db/(\tilde{\eps}\hbar)$ is a feedback parameter accounting for the TLA self-interaction mediated by the graphene sheet. Neglecting retardation and nonlocal effects, $\Gm_\ww(\rb,\rb)$ conveniently admits closed-form expressions that we use here to describe the atom-graphene interaction over a range of parameters (see SI for details). Incidentally, the dephasing $\gamma$ in Eqs.\ \eqref{eq:rho_ij_eom} is phenomenologically generalized to include additional possible decoherence channels, such that $\gamma\to\Gamma/2$ when relaxation occurs purely due to spontaneous emission.

The physics of Eqs.\ \eqref{eq:rho_ij_eom} has been extensively discussed in the context of atom-plasmon interactions occurring in optically-driven semiconductor quantum dot (SQD)-metal nanoparticle (MNP) hybrid systems \cite{artuso2008optical,artuso2010strongly,malyshev2011optical,malyshev2012condition,arrieta2014modelling}, and we briefly summarize the role of the parameters here: The TLA is driven at the effective Rabi frequency $\Omega$, accounting for the external field and its reflection from the nanophotonic element (here the graphene sheet); meanwhile, the dipole induced in the TLA $\db\rhot_{21}\ee^{-\ii\ww t}+\text{c.c.}$ produces a field that is reflected back on itself by the graphene sheet, resulting in a renormalization of the transition frequency $\Delta\to\Delta+\Ree\{G\}\ccpar{\rho_{22}-\rho_{11}}$ and dephasing rate $\gamma\to\gamma+\Imm\{G\}\ccpar{\rho_{22}-\rho_{11}}$. The self-interaction term $G$ thus endows the TLA response with an additional nonlinearity determined by geometric considerations (e.g., the separation distance) and the intrinsic optical properties of the nanophotonic environment, which are difficult to tune actively in SQD-MNP hybrids. The Lamb shift and Purcell enhancement introduce further sensitivity to changes in the local photonic density of states, but are often neglected in theoretical works describing similar systems. Here, through minor changes to the Fermi energy $\EF$, the graphene sheet enables electrical modulation of the parameter $G$, which mediates the driven TLA self-interaction and quantum electrodynamic effects, also (to a lesser extent) modifying the effective Rabi frequency $\Omega$, in the spectral neighborhood of the $2\EF$ interband transition, as shown in Fig.\ \ref{fig1}b for a dipole oriented parallel to the graphene sheet.

\begin{figure*}[t]
    \centering
    \includegraphics[width=\textwidth]{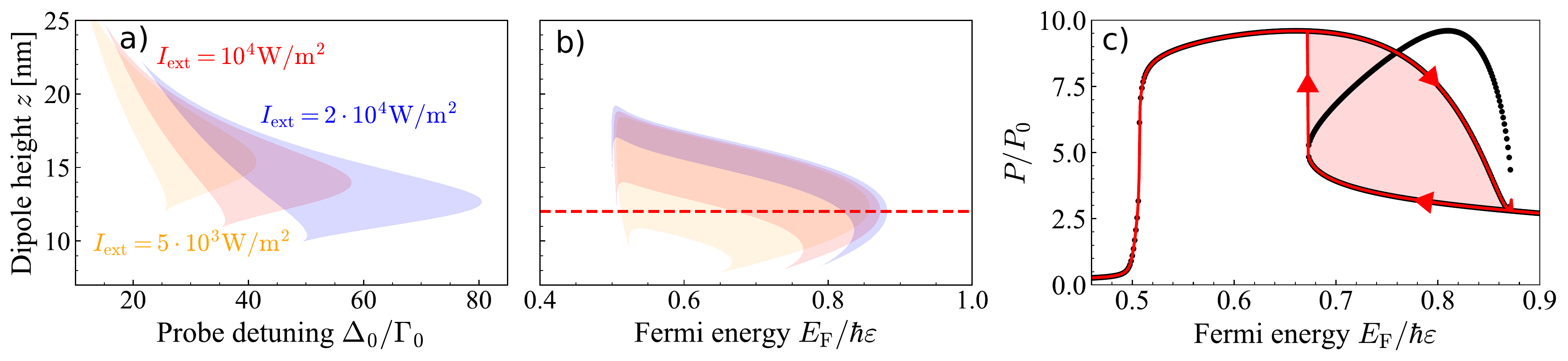}  
    \caption{\textbf{Regimes of optical bistability.} (a) Regions of bistability indicated by the discriminant of the third order polynomial governing the steady state of the two-level atom (TLA) for various impinging light intensities $I^{\rm ext}$ as a function of the probe detuning $\Delta_0$ and distance $z$ between the TLA and graphene sheet. (b) Same as in panel (a) but sweeping the Fermi energy $\EF$ at fixed detuning. (c) Normalized average radiation power as a function of $\EF$ for $z=12$\,nm, corresponding to the horizontal dashed line in panel (b). TLA and graphene parameters are the same as those in Fig.\ \ref{fig1} unless otherwise specified.}
\label{fig2}
\end{figure*}

Under steady-state conditions $\dot{\hat{\rho}}=0$, Eqs.\ \eqref{eq:rho_ij_eom} can be expressed as
\begin{subequations}
\begin{align}
    \frac{4\gamma}{\Gamma} |\Omega|^2 &= - \frac{Z+1}{Z} \sqpar{\left(\Delta - \Ree\{G\} Z\right)^2 + \left( \gamma - \Imm\{G\}Z \right)^2} , \label{eq:Z_SS} \\
    \tilde{\rho}_{21} &= \frac{\Omega Z}{(\Delta - \Ree\{G\}Z) + \ii(\gamma-\Imm\{G\}Z)} ,
\end{align}
\end{subequations}
where $Z\equiv\rho_{22}-\rho_{11}$ is the TLA population difference. The steady-state solution of the driven TLA is thus determined by Eq.\ \eqref{eq:Z_SS}, a third-order polynomial in $Z$ that admits up to three real solutions. More specifically, the nonlinearity introduced by the self-interaction or feedback parameter $G$ can render the TLA bistable when three real distinct solutions can be realized, such that the system admits two stable solutions and one unstable solution.

External control over optical bistability is explored in Fig.\ \ref{fig1}(c-e), where steady-state solutions from Eq.\ \eqref{eq:Z_SS} are plotted as black dots and superimposed red curves indicate direct time-domain solutions of Eqs.\ \eqref{eq:rho_ij_eom} in the steady state while adiabatically sweeping the impinging light intensity in Fig.\ \ref{fig1}(c), spectral detuning in Fig.\ \ref{fig1}(d), and graphene Fermi energy $\EF$ in Fig.\ \ref{fig1}(e). Importantly, time-domain solutions reveal hysteresis loops that are sensitive to the direction of change in the various external parameters considered, and thus access different bistable regimes of the TLA population. As the control parameters are varied, pairs of steady state solutions appear or disappear at fold bifurcation points, around which small variations in the control parameter can dramatically change the population difference, such that the bifurcations can be deemed ``catastrophic'' \cite{scheffer2009early,strogatz2018nonlinear}. Qualitatively similar optical bistability and hysteresis loops have been observed in experimental studies of dilute Rydberg gases when varying the intensity or frequency of the impinging laser field \cite{carr2013nonequilibrium}, while the graphene-atom system enables \textit{in situ} reversible optoelectronic tuning of the TLA state by varying the Fermi energy in graphene, in qualitative agreement with predictions for a driven TLA near an tunable indium tin oxide film \cite{arrieta2014modelling}.

Optoelectronic bistability in the TLA-graphene system is associated with a positive discriminant of the third-order polynomial in Eq.\ \eqref{eq:Z_SS}, which we map over the TLA-graphene separation $z$ at several impinging light intensities in Figs.\ \ref{fig2}(a,b) while varying detuning and Fermi energy, respectively. In the former situation, the light intensity primarily shifts the spectral window where bistability emerges, while in the latter case the intensity affects the range of separations. Linear stability analysis (see SI for details) reveals that the stability of the steady-state solutions in Figs.\ \ref{fig2}(a,b) are indeed comprised of two stable solutions and one unstable solution. The bistable state of the coupled TLA-graphene system can be observed in the total normalized radiation power from the TLA, obtained by integrating the dipole radiation pattern over all angles \cite{novotny2012principles}, and presented in Fig.\ \ref{fig2}(c) at a separation distance of $z=12$\,nm and detuning $\hbar\Delta_0=8$\,$\mu$eV corresponding to the horizontal dashed line in Fig.\ \ref{fig2}(b). Notably, by tuning $\EF$ in graphene, the radiation power can achieve a ten-fold enhancement for the parameters under consideration, while the hysteresis behavior presents a clear signature of atomic bistability.

\begin{figure}
    \centering
    \includegraphics[width=\linewidth]{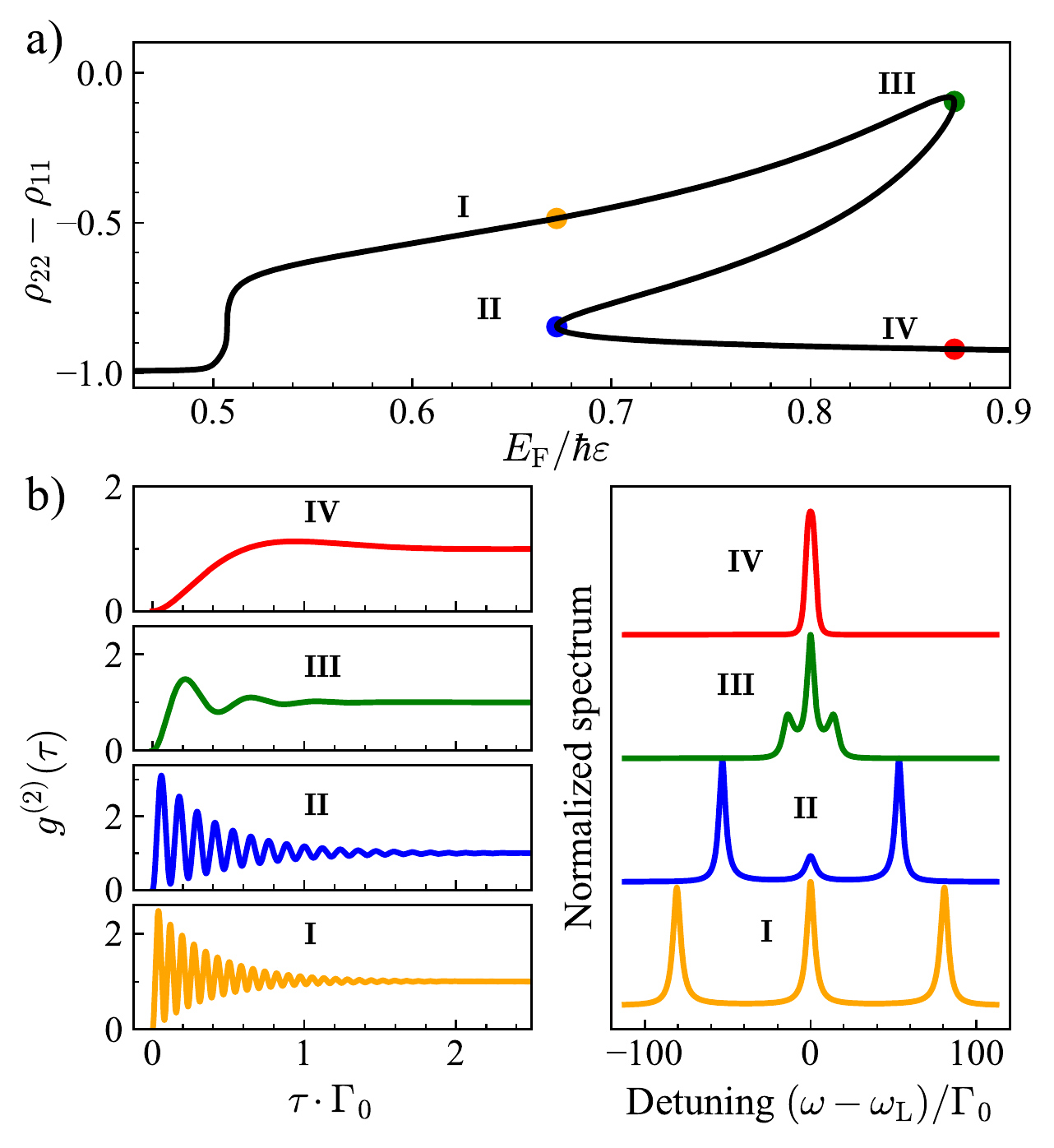}  
    \caption{{\bf Signatures of optical bistability in resonance fluorescence and antibunching.} (a) Steady-state solutions of the TLA population difference $Z$ obtained from Eq.\ \eqref{eq:Z_SS} are plotted as a function of the graphene Fermi energy $\EF$ when the TLA is $z=12$\,nm above the graphene sheet and driven at a frequency corresponding to $\Delta=8$\,$\mu$eV with a field of intensity $I^{\rm ext}=10^4$\,W/m$^2$. (b) The second-order correlation function $g^{(2)}(\tau)$ (left panel) and resonance fluorescence spectra (right panel) are presented as functions of delay time $\tau$ and  detuning $\omega-\omega_{\rm L}$, respectively, for system parameters indicated by the color-coordinated points I-IV in panel (a).}
    \label{fig3}
\end{figure}

Atomic bistability also manifests in the spectrum of the fluorescent light emitted from the TLA, obtained from the first-order correlation function of the emitted field \cite{meystre2007elements}. Typically, for weak excitation, incident and fluorescent light frequencies coincide, and the incoherent spectrum forms a single Rayleigh peak; under strong excitation, the TLA spectrum exhibits a large central peak and two sidebands---the so-called Mollow triplet \cite{mollow1969power}. As the upper and lower TLA population branches are accessed by tuning the Fermi energy in Fig.\ \ref{fig3}(a), the fluorescence spectrum in the right panel of Fig.\ \ref{fig3}(b) exhibits a bistable response, with states distinguished by variations spanning a single Rayleigh peak to the Mollow triplet at the indicated points. Note that for an isolated atom, sidebands split continuously from the Rayleigh peak as the incident field intensity increases, in a manner roughly equivalent to a second-order phase transition \cite{bonifacio1976cooperative, bonifacio1978optical}. In contrast, the discontinuous change in the fluorescence spectrum exhibited by the TLA-graphene system when tuning $\EF$ (e.g., between regions I and II or III and IV) is reminiscent of first-order phase transitions in a system at thermal equilibrium.

The second-order correlation function $g^{(2)}(\tau)$ associated with the emitted light at the selected points is presented in the left panel of Fig.\ \ref{fig3}(b), presenting antibunching for vanishing time delay $\tau=0$ while approaching unity as $\tau\rightarrow\infty$ \cite{meystre2007elements,mohammadzadeh2019resonance}, either monotonically evolving or rapidly oscillating between these limits depending on the region of bistability. Analogous to the resonance fluorescence spectrum, the transient statistics of the emitted light can be abruptly modified by traversing the hysteresis curve (e.g., from region III to IV).

\begin{figure*}
    \centering
    \includegraphics[width=\textwidth]{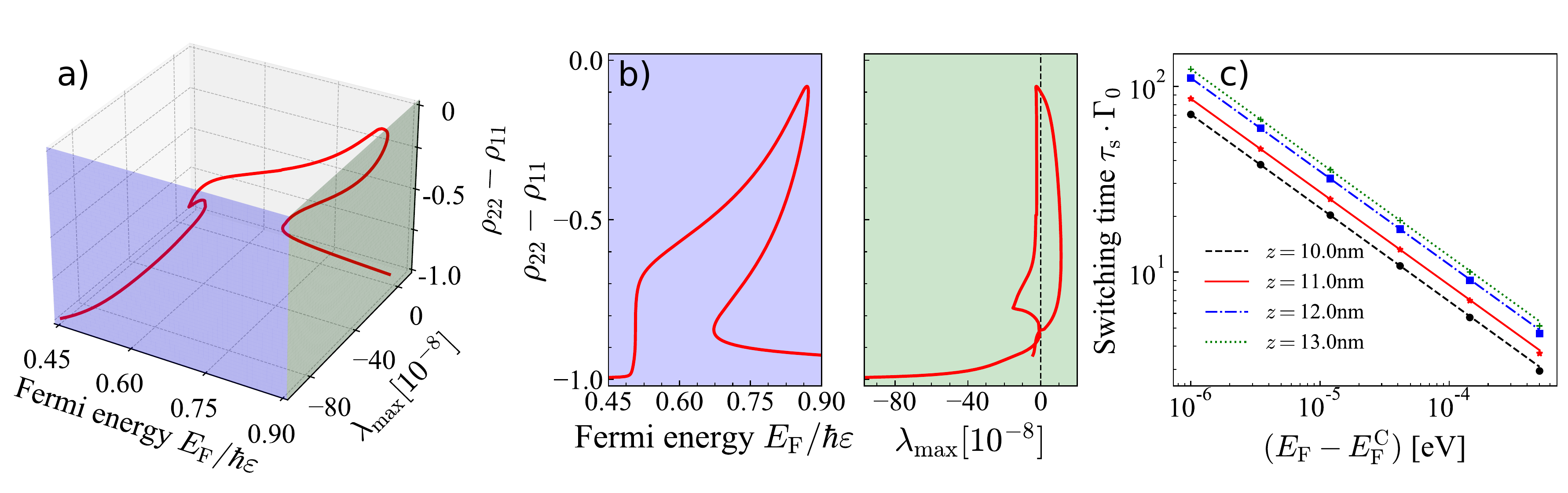}
    \caption{\textbf{Critical slowing down near bifurcations.} (a) The population difference as a function of the Fermi energy and maximal eigenvalue of the system Jacobian. (b) The 3D-plot in panel (a) is projected on indicated planes to more clearly show the zeros of the maximum eigenvalue. (c) Critical slowing down as a function of the $\EF$ above the critical point, as explained in the text for different distances $z$. The parameters used here are the same as those in Fig. \ref{fig2}.}
    \label{fig4}
\end{figure*}

The characteristic timescale $\tau_{\rm s}$ on which a TLA in a bistable state can be brought from one stable branch to another is an important metric for optical switching applications \cite{nugroho2013bistable,carreno2018nonlinear}, and has been exploited to herald phase transitions in dilute Rydberg ensembles \cite{carr2013nonequilibrium}. While approaching a bifurcation point (see vertical arrows in Fig.\ \ref{fig1}), the system becomes increasingly slow at recovering from perturbations \cite{scheffer2009early}. The slow-down of a bistable device at such critical points limits its response time but does not severely inhibit functionality, since the phase boundary parameter (in this case $\EF$) can be brought to larger values, taking advantage of the power law scaling of critical slow-down \cite{bonifacio1979critical}. Regions of critical slow-down are revealed by analyzing the Jacobian characterizing the perturbed system in a linear stability analysis: when the Jacobian eigenvalues are all negative, the system is considered stable against small perturbations, and otherwise is exponentially unstable \cite{lugiato2015nonlinear}. In Fig.\ \ref{fig4}(a) we present the maximum real part of these eigenvalues as a function of $\EF$. The system exhibits critical slow-down when the maximum eigenvalue changes sign, which occurs at two regions in Fig.\ \ref{fig4}(a), namely at the border of the upper branch and moving from the lower to upper branch.

When tuning a control parameter (e.g., Fermi energy, intensity, or detuning) to trigger a transition from the lower to upper stable branch along the hysteresis curve, i.e., from point IV to I in Fig.\ \ref{fig3}, critical slow-down occurs at II. The critical slow-down is characterized by a power law in the control parameter, e.g., $\tau_{\rm s}\propto |\EF - E_{\rm F}^{\rm c}|^{-\alpha}$, where $\tau_{\rm s}$ is defined here as the time until maximum population inversion is reached (as indicated in the SI) \cite{nugroho2013bistable}, $E_{\rm F}^{\rm c}$ is the critical Fermi energy, and $\alpha$ is the critical exponent. In Fig.\ \ref{fig4}c, we plot $\tau_{\rm s}$ as as $\EF$ is tuned for various TLA separation distances $z$. Interestingly, the critical exponents $\alpha$ are independent of $z$, as is also the case for other control parameters (not shown here), indicating that the critical exponent $\alpha \approx 0.5$ characterizing the bistable transition is universal; these findings are corroborated by reports of cooperative interactions among quantum emitters, e.g., in a theoretical study of an optically-driven TLA coupled with a plasmonic nanoparticle \cite{nugroho2013bistable}, or in measurements of a dilute Rydberg gas ensemble \cite{carr2013nonequilibrium}. Note that, while we have defined $\tau_{\rm s}$ as the maximum of $Z$ in a transient regime, the true power law---obtained precisely at the critical point---is expected to exhibit small quantitative deviations.

In summary, we propose to harness the interband transition threshold in graphene to achieve electro-optical control of atomic bistability. Here, optical bistability emerges from the self-interaction of the atomic transition dipole in the presence of the graphene sheet, which also influences the Purcell effect and Lamb shift, leading to a rich interplay between the driven optical nonlinearity and quantum electrodynamic response as the Fermi energy is modulated. We show that the bistability and hysteresis behavior resulting from the atom-graphene interaction can be observed in the light scattered by the atom, specifically in the radiated power, resonance fluorescence spectrum, and photon statistics. In particular, electrical tuning of the graphene sheet can trigger critical slow-down of the steady-state approach in the two-level atom dynamics, which is heralded in the scattering spectrum by a discontinuous quench of the sidebands or Rayleigh peak in the Mollow triplet when traversing hysteresis loops. The scheme presented here to electrically tune the optically-driven atom-graphene interaction can be used to coherently control quantum states for explorations in atomic physics (e.g., combined with optical lattices), and constitutes a versatile platform for quantum nano-optics.


\section{Acknowledgements}

The authors thank Fabio Raspanti, P.~A.~D.~Gon\c{c}alves, Sebastian Hofferberth, N.~Asger Mortensen, and Klaus M{\o}lmer for insightful and enjoyable discussions.
The Center for Nano Optics is financially supported by the University of Southern Denmark (SDU 2020 funding).
J.~D.~C. is a Sapere Aude research leader supported by Independent Research Fund Denmark (grant no. 0165-00051B).


%

\end{document}


\title{Optoelectronic control of atomic bistability with graphene  \\ {\color{gray} \small -- SUPPLEMENTARY INFORMATION --}}

\author{Mikkel~Have~Eriksen}
\affiliation{Center for Nano Optics, University of Southern Denmark, Campusvej 55, DK-5230 Odense M, Denmark}

\author{Jakob~E.~Olsen}
\affiliation{Faculty of Engineering, University of Southern Denmark, Campusvej 55, DK-5230 Odense M, Denmark}

\author{Christian~Wolff}
\affiliation{Center for Nano Optics, University of Southern Denmark, Campusvej 55, DK-5230 Odense M, Denmark}

\author{Joel~D.~Cox}
\email[Joel~D.~Cox: ]{cox@mci.sdu.dk}
\affiliation{Center for Nano Optics, University of Southern Denmark, Campusvej 55, DK-5230 Odense M, Denmark}
\affiliation{Danish Institute for Advanced Study, University of Southern Denmark, Campusvej 55, DK-5230 Odense M, Denmark}

\begin{abstract}
We review the formalism describing a two-level atom driven by a classical optical field and interacting with electromagnetic vacuum fluctuations mediated by an arbitrary nanophotonic geometry. We then introduce the Green's tensor characterizing the interaction of a point dipole with an isotropic 2D material, which admits closed-form expressions in the quasistatic approximation that are used to compute the Lamb shift and spontaneous emission rate enhancement of an emitter near a graphene sheet. Subsequently, we present the prescription to quantify and assess the stability of steady state solutions to the optical Bloch equations, and outline the procedures used to compute the resonance fluorescence spectrum and second-order correlation function of the two-level system driven by a monochromatic classical field in a nanophotonic environment. Finally, we discuss the convention used to assess critical slow-down of the bistable system and determine the associated critical exponent.
\end{abstract}

\date{\today}
\maketitle
\tableofcontents

\section{Master equation formalism}

It is instructive to revisit the quintessential quantum optics problem of a two-level atom (TLA) driven by a classical electromagnetic field and coupling with electromagnetic vacuum fluctuations: The TLA itself is characterized by the Hamiltonian
\begin{equation} \label{eq:H_A}
    \Hm_{\rm A} = \hbar\sum_{j=1}^2\vep_j\ket{j}\bra{j} ,
\end{equation}
where $\hbar\vep_j$ are the energies associated with the stationary states $\ket{j}$ forming a complete basis for $j\in\{1,2\}$; the interaction of the atom with a classical monochromatic field $\Eb(\rb,\ww_{\rm L})\ee^{-\ii\ww_{\rm L}t} + {\rm c.c.}$ is governed by
\begin{equation} \label{eq:H_AL}
    \Hm_{\rm AL} = -\db\cdot\sqpar{\Eb(\rb,\ww_{\rm L})\ee^{-\ii\ww_{\rm L}t} + {\rm c.c.}}\ccpar{\ket{1}\bra{2}+\ket{2}\bra{1}} ,
\end{equation}
where $\rb=(x,y,z)$ is the position of the TLA and $\db\equiv\braket{1|\hat{\db}|2}=\braket{2|\hat{\db}|1}$ is the transition matrix element of the dipole operator $\hat{\db}$; fluctuations in the vacuum electromagnetic reservoir are described by the Hamiltonian
\begin{equation} \label{eq:H_R}
    \Hm_{\rm R} = \hbar\int_0^\infty d\ww \ww \int d^3\rb \fh_\ww^\dagger(\rb)\cdot\fh_\ww(\rb) ,
\end{equation}
where $\fh^\dagger$ ($\fh$) are bosonic field creation (annihilation) operators; finally, the Hamiltonian governing the interaction of the atom with the reservoir is
\begin{equation} \label{eq:H_AR}
    \Hm_{\rm AR} = -\int_0^\infty d\ww \db\cdot\sqpar{\hat{\Eb}_{\rm R}(\rb,\ww) + {\rm h.c.}} \ccpar{\ket{1}\bra{2}+\ket{2}\bra{1}} ,
\end{equation}
where
\begin{equation}
    \hat{\Eb}_{\rm R}(\rb,\ww) = \ii\sqrt{\frac{\hbar}{\pi\eps_0}}\frac{\ww^2}{c^2}\int d^3\rb' \sqrt{\Imm\{\chi_\ww(\rb')\}} \Gm_\ww(\rb,\rb')\cdot\fh_\ww(\rb')
\end{equation}
is the electromagnetic field associated with vacuum fluctuations, constructed from the dielectric susceptibility $\chi_\ww$ and the classical electromagnetic Green's dyadic $\Gm_\ww$ \cite{scheel2008macroscopic,forati2014graphene}. Identifying the bare ($\Hm_{\rm A}$ and $\Hm_{\rm R}$) and interaction ($\Hm_{\rm AL}$ and $\Hm_{\rm AR}$) terms of the total Hamiltonian, we may transform a generic operator $\mathcal{A}$ in the Schr\"odinger picture to the interaction picture as $\mathcal{A}\to\ee^{\ii(\Hm_{\rm A}+\Hm_{\rm R}) t /\hbar}\mathcal{A}\ee^{-\ii(\Hm_{\rm A}+\Hm_{\rm R})t /\hbar}$.

Here we are interested in the evolution of the atomic system characterized by only two states, while the photonic reservoir spans infinite degrees of freedom. We may thus define a ``joint'' density matrix operator $\rhoSR$, where the subscripts S and R indicate states of the atomic system and photonic reservoir, respectively; in the absence of any interaction between the system and the reservoir (i.e., taking $\VAR\to0$), the joint density matrix operator can be factorized as $\rhoSR(t)=\rhoS(t)\otimes\rhoR(t_0)$, where $t_0$ is a time at which the reservoir is in equilibrium. If the system-reservoir interaction is weak, so that any associated changes to the reservoir are negligibly small, we can write
\begin{equation}
    \rhoSR(t) = \rhoS(t)\otimes\rhoR(t_0) + \delta\rho(t) ,
\end{equation}
where $\delta\rho(t)$ is a fluctuation due to the system-reservoir interaction. In this formalism, the dynamics of the atomic system are governed by the \textit{reduced} density matrix operator $\rhoS\equiv\TrR\{\rhoSR\}$, obtained by tracing over the reservoir degrees of freedom to ensure that $\TrR\{\delta\rho(t)\}=0$.

In the interaction representation, the density matrix equation of motion is
\begin{equation} \label{eq:rho_SR_eom}
    \pd{\rhoSR}{t} = -\frac{\ii}{\hbar} \sqpar{\VAL + \VAR, \rhoSR} ,
\end{equation}
where the Hamiltonian terms
\begin{equation} \label{eq:V_AL}
    \VAL = - \sqpar{\db\cdot\Eb(\rb,\wL)\ee^{-\ii\wL t} + {\rm c.c.}}\ccpar{\ket{1}\bra{2}\ee^{-\ii\vep t} + {\rm h.c.}}
\end{equation}
and 
\begin{equation} \label{eq:V_AR}
    \VAR = - \sqrt{\frac{\hbar}{\pi\eps_0}} \int_0^\infty d\ww \frac{\ww^2}{c^2} \int d^3\rb \sqpar{\ii\sqrt{\Imm\{\chi_\ww(\rb)\}}\db\cdot\Gm_\ww(\rb,\rb')\cdot\fh_\ww(\rb')\ee^{-\ii\ww t} + {\rm h.c.}} \ccpar{\ket{1}\bra{2}\ee^{-\ii\vep t} + {\rm h.c.}},
\end{equation}
obtained by transforming Eqs.\ \eqref{eq:H_AL} and \eqref{eq:H_AR} to the interaction picture, govern the atom-light and atom-reservoir interaction, respectively, and are expressed in terms of the TLA transition frequency $\vep \equiv\vep_2-\vep_1$. By formally integrating Eq.\ \eqref{eq:rho_SR_eom}, we obtain
\begin{equation} \label{eq:rho_SR_integrated}
    \rhoSR(t) = \rhoSR(0) - \frac{\ii}{\hbar}\int_0^t dt' \sqpar{\VAL(t') + \VAR(t'), \rhoSR(t')} ,
\end{equation}
where $t=0$ designates the time at which the interaction $\Vm_{\rm AR}$ ``starts''. Inserting the solution Eq.\ \eqref{eq:rho_SR_integrated} into Eq.\ \eqref{eq:rho_SR_eom} and tracing over reservoir states thus leads to
\begin{equation} \label{eq:rho_S_eom}
    \pd{\rhoS}{t} = -\frac{\ii}{\hbar}\sqpar{\VAL,\rhoS} - \frac{1}{\hbar^2}\TrR\clpar{\int_0^t dt' \sqpar{\VAR(t),\sqpar{\VAR(t'),\rhoSR(t')}}} ,
\end{equation}
where the terms of linear order in $\VAR$ vanish in the trace since $\av{\fh}=\av{\fh^\dagger}=0$. In the framework of Markovian dynamics, the second term in Eq.\ \eqref{eq:rho_S_eom} can be evaluated by assuming that the density matrix is independent of its history, i.e., $\rho(t')\approx\rho(t)$, and that the integrand decays quickly enough towards zero that the upper integration boundary can be extended towards infinity; in the frequency domain, the Markov condition is valid when the spectral variation of the photonic reservoir is negligible on frequency scales $\sim\Gamma_0$. In performing the trace over reservoir states, we assume that the reservoir is in a thermal equilibrium such that $\av{\fh_\ww^\dagger(\rb)\fh_{\ww'}(\rb')} = N(\ww)\delta(\rb-\rb')\delta(\ww-\ww')$ \cite{scully1999quantum, antao2021two}, where $N(\ww)=\left(\ee^{\hbar\ww/\kB T}-1\right)^{-1}$ is the Bose-Einstein distribution, and other permutations are readily obtained from the commutation relation $\sqpar{\fh_\ww(\rb),\fh_{\ww'}^\dagger(\rb')}=\delta(\rb-\rb')\delta(\ww-\ww')$. Then, simplifying the notation by replacing $\rhoS\to\rho$, we obtain the master equation for the density matrix
\begin{equation}
    \pd{\rho}{t} = -\frac{\ii}{\hbar}\sqpar{\VAL + \hbar\delta\ww\ket{2}\bra{2}, \rho} + \frac{\Gamma}{2}\ccpar{2\ket{1}\bra{2}\rho\ket{2}\bra{1}-\ket{2}\bra{2}\rho-\rho\ket{2}\bra{2}} ,
\end{equation}
where
\begin{equation} \label{eq:Lamb}
    \delta\ww = \frac{\mu_0}{\pi\hbar}\mathcal{P}\int_0^\infty d\ww\frac{\ww^2}{\vep-\ww}\Imm\clpar{\db\cdot\Gm_{\ww}(\rb,\rb)\cdot\db}
\end{equation}
denotes the Lamb shift as a principle value integral and
\begin{equation} \label{eq:Decay_rate_enhancement}
    \Gamma = \frac{2\mu_0}{\hbar}\vep^2\Imm\clpar{\db\cdot\Gm_\vep(\rb,\rb)\cdot\db} 
\end{equation}
is the spontaneous emission rate of the TLA. The integral of Eq.\ \eqref{eq:Lamb} diverges due to the free-space part of the Green's function; while the divergence can be dealt with using a renormalization, we adopt the common approach of assuming that the free-space contribution of the Green's function is already incorporated in the atomic transition frequency $\vep$. Then, only the reflected part of the Green's function $\Gm_\ww^{\rm ref}$ contributes to Eq.\ \eqref{eq:Lamb}, which can be approximated using the Kramer-Kronigs relations as
\begin{equation} \label{eq:Lamb_shift_approx}
    \delta\ww = -\frac{\mu_0}{\hbar}\vep^2\Ree\clpar{\db\cdot\Gm_\vep^{\rm ref}(\rb,\rb)\cdot\db} .
\end{equation}
The above approximation assumes that only frequencies near the transition frequency $\vep$ contribute to the Lamb shift \cite{dzsotjan2011dipole}, and is shown in Fig.\ \ref{fig:Lamb_comparison} to describe the same behavior as Eq.\ \eqref{eq:Lamb} for electrically-tuned graphene.

\section{Electrodynamics of a dipole interacting with a graphene sheet}

We consider a point dipole with moment $\db$ located at a point $\rb'=(x',y',z')$ near an interface defined in the $(x,y,z_0)$ plane that is characterized by Fresnel reflection coefficients $\rs$ and $\rp$ for s- and p-polarized fields, respectively. The field produced by the dipole is then given from
\begin{equation}
    \Eb(\rb,\ww) = \mu\ww^2\sqpar{\Gm_\ww^{(0)}(\rb,\rb') + \Gm_\ww^{\rm ref}(\rb,\rb')}\cdot\db,
\end{equation}
where $\Gm_\ww^{(0)}$ and $\Gm_\ww^{\rm ref}$ are the bare and reflected parts of the electromagnetic Green's tensor. Following Refs.\ \cite{novotny2012principles} and \cite{hohenester2020nano}, the reflected part of the Green's tensor at the dipole location (i.e., for $\rb=\rb'$) is expressed as
\begin{equation} \label{eq:G_ref_rr}
    \Gm_\ww^{\rm ref} = \frac{\ii}{8\pi k_{\rm a}^2}\int^\infty_0d\kpar\kpar\frac{\ee^{\ii 2k_{{\rm a},z} z}}{k_{{\rm a},z}} \begin{bmatrix} k_{\rm a}^2 \rs(\kpar) - k_{{\rm a},z}^2 \rp(\kpar) & 0 & 0 \\ 0 & k_{\rm a}^2 \rs(\kpar) - k_{{\rm a},z}^2 \rp(\kpar) & 0 \\ 0 & 0 & 2 \kpar^2 \rp(\kpar) \end{bmatrix} ,
\end{equation}
where $k_{m,z}=\sqrt{\eps_m(\ww^2/c^2)-\kpar^2 + \ii 0^+}$ denotes the normal component of the wave vector $\kb_m$ in a medium with relative permittivity $\eps_m$ and $\kpar=\sqrt{k_x^2+k_y^2}$ is the conserved in-plane component, while $m={\rm a}$ indicates the medium above the interface. Note that the result of Eq.\ \eqref{eq:G_ref_rr} is obtained by assuming that the reflection coefficients are isotropic in the plane of the interface. For a two-dimensional (2D) material characterized by an isotropic optical 2D conductivity $\sigma(\kpar,\ww)$, the reflection coefficients associated with impinging s- and p-polarized fields are \cite{gonccalves2020plasmonics}
\begin{subequations} \label{eq:Fresnel}
\begin{align}
    \rs &= \frac{k_{{\rm a},z} - k_{{\rm b},z} - \mu_0\ww\sigma(\kpar,\ww)}{k_{{\rm a},z} + k_{{\rm b},z} + \mu_0\ww\sigma(\kpar,\ww)}  \\
    \rp &= \frac{\epsb k_{{\rm a},z} - \epsa k_{{\rm b},z} + k_{{\rm a},z}k_{{\rm b},z}\sigma(\kpar,\ww)/(\ww\eps_0)}{\epsb k_{{\rm a},z} + \epsa k_{{\rm b},z} + k_{{\rm a},z}k_{{\rm b},z}\sigma(\kpar,\ww)/(\ww\eps_0)} ,
\end{align}
\end{subequations}
where $m={\rm a}$ and $m={\rm b}$ indicate media directly above and below the 2D material, respectively.

To simplify the integration in Eq.\ \eqref{eq:G_ref_rr}, we assume that the dipole is sufficiently close to the 2D material so that retardation effects can be safely neglected, which amounts to writing
\begin{equation}
    \lim_{c\to\infty} k_{m,z} = \ii \sqrt{k_x^2+k_y^2} \equiv \ii Q .
\end{equation}
Eq.\ \eqref{eq:G_ref_rr} then becomes
\begin{equation} \label{eq:G_ref_Q}
    \Gm_\ww^{\rm ref} = \frac{1}{8\pi k_{\rm a}^2}\int^\infty_0dQ \ee^{-2Qz} \begin{bmatrix} k_{\rm a}^2 \rs(Q) + Q^2 \rp(Q) & 0 & 0 \\ 0 & k_{\rm a}^2 \rs(Q) + Q^2 \rp(Q) & 0 \\ 0 & 0 & 2 Q^2 \rp(Q) \end{bmatrix} ,
\end{equation}
where the reflection coefficients of Eq.\ \eqref{eq:Fresnel} are now
\begin{subequations} \label{eq:Fresnel_Q}
\begin{align}
    \rs &\approx -\frac{\mu_0\ww\sigma(Q,\ww)}{\mu_0\ww\sigma(Q,\ww) + 2\ii Q}  \\
    \rp &\approx \frac{(\epsb - \epsa)\eps_0\ww + \ii Q \sigma(Q,\ww)}{(\epsb + \epsa)\eps_0\ww + \ii Q \sigma(Q,\ww)} .
\end{align}
\end{subequations}
We remark that the quasistatic approximation employed here has been found to be well-justified for the dipole-graphene separation distances considered in this work, as reported in Ref.\ \cite{koppens2011graphene}.

To further simplify calculations, we may neglect nonlocal effects in the 2D conductivity by replacing $\sigma(Q,\ww)\rightarrow\sigma(\ww)\equiv\lim_{Q\to0}\sigma(Q,\ww)$, where the local conductivity $\sigma(\ww)$ is independent of the in-plane wave vector $Q$, so that Eq.\ \eqref{eq:G_ref_Q} admits the closed-form expressions
\begin{subequations}
\begin{equation} \label{eq:G_ref_par}
    \Gm^{\rm ref}_{\ww,\parallel} = \frac{\zeta}{8\pi}\ee^{-2\zeta z}E_1(-2\zeta z) + \frac{1}{2}\Gm^{\rm ref}_{\ww,\perp} 
\end{equation}
and
\begin{equation} \label{eq:G_ref_perp}
    \Gm^{\rm ref}_{\ww,\perp} = \frac{c^2}{4\pi\epsa\ww^2}\clpar{\frac{2\ii\epsa\eps_0\ww}{\sigma(\ww)}\sqpar{\frac{1}{4z^2} + \frac{\eta}{2z} + \eta^2\ee^{-2\eta z}E_1(-2\eta z)} + \frac{1}{4z^3}}
\end{equation}
\end{subequations}
for dipoles oriented parallel and perpendicular to the 2D sheet, respectively, where the parameters $\zeta \equiv \ii\mu_0\ww\sigma(\ww)/2$ and $\eta \equiv \ii(\epsa + \epsb)\eps_0\ww/\sigma(\ww)$ are introduced to simplify the notation and $E_1(z)=\int^\infty_z dt \ee^{-t}/t$ is the exponential integral for a complex argument $z$.
    
\subsection{Optical response of graphene}

\begin{figure} \label{fig:decayrate_nonlocal}
    \centering
    \includegraphics[width=0.85\textwidth]{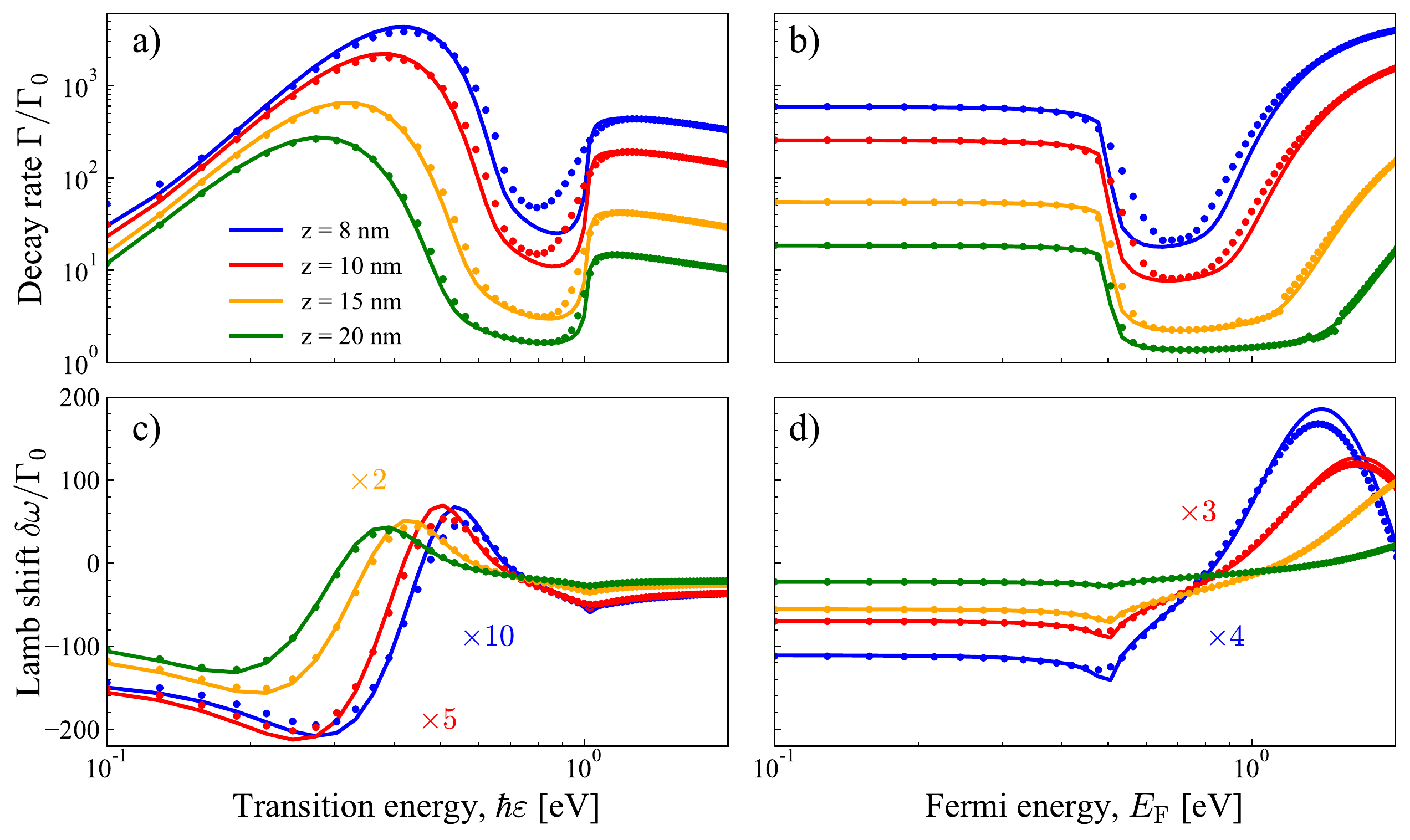}
    \caption{{\bf Purcell factor and Lamb shift.} The emission rate $\Gamma$ of a dipole near a graphene sheet is presented as a function of (a) the transition energy $\hbar\vep$ and (b) the graphene Fermi energy $\EF$ for the various graphene-dipole separations $z$ indicated in the legend of (a), where solid and dotted curves correspond to local and nonlocal graphene conductivity models, respectively; the analogous Lamb shift $\delta\ww$ is presented in panels (c) and (d). The Purcell factor and Lamb shift are normalized to the vacuum spontaneous emission rate $\Gamma_0=0.38$\,ns$^{-1}$, and we consider $\EF = 0.51$\,eV in panels (a,c) and $\hbar\vep = 1$\,eV in panels (b,d).}
\end{figure}

Turning to the specific case of graphene, the conductivity is obtained from the well-known expression \cite{koppens2011graphene}
\begin{equation} \label{eq:sigma_k}
    \sigma(\kpar,\ww) = \frac{-\ii\ww(1+\ii/\ww\tau)\chi(\kpar,\ww+\ii/\tau)}{1+(\ii/\ww\tau)\chi(\kpar,\ww+\ii/\tau)/\chi(\kpar,0)} ,
\end{equation}
where inelastic scattering at a phenomenological rate $\tau^{-1}$ is introduced following the prescription of Mermin to conserve electron population \cite{mermin1970lindhard}, and the linear response function $\chi(\kpar,\ww)$ is obtained in the zero-temperature limit of the random-phase approximation (RPA) as \cite{koppens2011graphene}
\begin{equation}
    \chi(\kpar,\ww) = \frac{e^2}{4\pi \hbar} \clpar{ \frac{8\kF}{\vF\kpar^2} + \frac{G(-\Delta_-) \Theta(- \text{Re}\{\Delta_-\} -1) + \sqpar{G(\Delta_- ) + \ii \pi} \Theta(\text{Re}\{ \Delta_-\} + 1) - G(\Delta_+)}{\sqrt{\ww^2 - \vF^2\kpar^2}} } ,
\end{equation}
with $G(z) = z\sqrt{z^2-1} - \log(z+\sqrt{z^2-1})$ and $\Delta_{\pm} = (\ww / v_{\rm F} \pm 2\kF ) / \kpar$, while the imaginary part of the logarithm is in the $(-\pi , \pi ]$ range and the square roots are chosen to yield positive real parts. In the limit of $\kpar\to0$, the local RPA conductivity of graphene is
\begin{equation} \label{eq:sigma_0}
    \sigma(\ww) =  \frac{e^2}{\pi\hbar^2}\frac{\ii\EF}{\ww+\itau}+\frac{e^2}{4\hbar}\sqpar{1+\frac{\ii}{\pi}\log\ccpar{\frac{\ww+\itau-2\EF/\hbar}{\ww+\itau+2\EF/\hbar}}} .
\end{equation}

\begin{figure}[t] \label{fig:Lamb_comparison}
    \centering
    \includegraphics[width=0.45\textwidth]{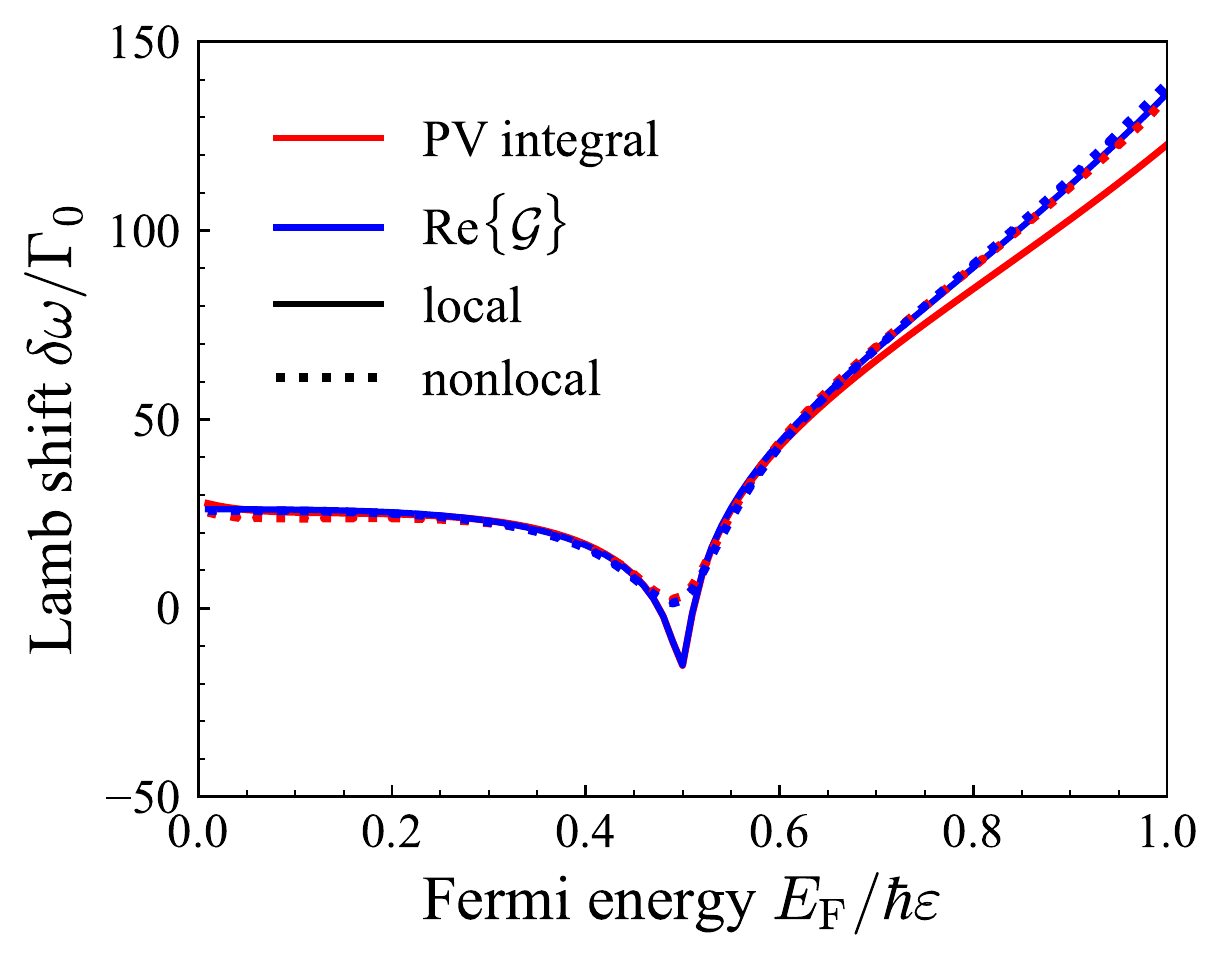}
    \caption{{\bf Lamb shift calculation.} The Lamb shift calculated by evaluating the principle value integral in Eq.\ \eqref{eq:Lamb} (PV integral, red curves) is contrasted with that predicted by the approximation of Eq.\ \eqref{eq:Lamb_shift_approx} ($\Ree\{\Gm\}$, blue curves), where the latter result is offset by $35$\,$\mu$eV.}
\end{figure}

In Fig.\ \ref{fig:decayrate_nonlocal}, we present the spontaneous emission rate of Eq.\ \eqref{eq:Decay_rate_enhancement} and the Lamb shift of Eq.\ \eqref{eq:Lamb_shift_approx} for a dipole placed at various distances $z$ from a graphene sheet, where calculations obtained in the local limit of the RPA, i.e., from Eqs.\ \eqref{eq:G_ref_par} and \eqref{eq:sigma_0} (solid curves), are contrasted with those incorporating nonlocal effects in the conductivity by combining Eqs.\ \eqref{eq:G_ref_rr} and Eq.\ \eqref{eq:sigma_k} (dotted curves). Results for the Purcell factor $\Gamma/\Gamma_0$ in Fig.\ \ref{fig:decayrate_nonlocal}(a,b) are in quantitative agreement with those presented in Ref.\ \cite{koppens2011graphene}, where nonlocal effects mainly soften the sharp features associated with the interband transition threshold near $\hbar\ww=2\EF$; similar smoothing due to nonlocal effects manifests in the Lamb shift $\delta\ww/\Gamma_0$ presented in Fig.\ \ref{fig:decayrate_nonlocal}(c,d), which also closely resembles analogous results presented in Ref.\ \cite{chang2017constructing} obtained by directly evaluating the principle value integral of Eq.\ \eqref{eq:Lamb}. To explore the validity of the approximation in Eq.\ \eqref{eq:Lamb_shift_approx}, we compare the Lamb shifts predicted by Eqs.\ \eqref{eq:Lamb} and \eqref{eq:Lamb_shift_approx} in Fig.\ \ref{fig:Lamb_comparison} when adopting either local (solid curves) or nonlocal (dotted curves) conductivities. The main difference between the approximate and exact Lamb shift calculation amounts to a constant $\sim35$\,$\mu$eV offset at $z=12$\, nm that is independent of Fermi energy. The discrepancy is presumably a result of the Dirac approximation of linearized electronic dispersion employed to treat the optical response of graphene, which leads to constant interband absorption at high frequencies, up to the $\ww\to\infty$ contour on which the Green's function is assumed to vanish in Eq.\ \eqref{eq:Lamb}. However, because a true optical response function would indeed vanish in the limit of infinite frequency, it is reasonable to expect that the rotating wave approximation used to obtain Eq.\ \eqref{eq:Lamb_shift_approx} actually provides a more realistic estimate of the Lamb shift when the optical response of graphene is described by conical electronic bands. We further point out that any constant energy offset between the models can be absorbed in the vacuum Lamb shift, which is already assumed to enter the TLA transition frequency $\vep$.

We additionally remark that the results of Fig.\ \ref{fig:decayrate_nonlocal} indicate that the spectral variation of the Green's function describing the photonic reservoir occurs on energy scales $\sim \hbar\tau^{-1}$, i.e., the inelastic scattering linewidth of graphene entering the 2D conductivity. Thus, the Markov approximation invoked in the master equation formalism of the previous section is reasonable when the intrinsic decay rate of the TLA satisfies $\Gamma_0\ll\tau^{-1}$. In particular, realistic incoherent electron scattering rates for graphene are on the order of $\gtrsim1$\,meV, while broadening associated with finite temperature effects occurs on the order of $\kB T\approx$25\,meV at room temperature. If we conservatively define a non-dispersive reservoir as varying $<10\%$ on an energy scale of the TLA linewidth, the TLA dynamics are Markovian when $\Gamma_0\leq \kB T/100\sim 250$\,$\mu$eV, while contributions from inelastic electron scattering and nonlocal effects will further increase the threshold value. In the main text, we consider intrinsic atomic linewidths $\hbar\Gamma_0$ on the order of $\mu$eV, so that the approximation is well-justified; a two-level system with a larger transition dipole moment may however trigger non-Markovian dynamics, e.g., non-Lorentzian resonances, as previously reported in the radiation dynamics of photonic crystals \cite{hoeppe2012direct}.

\section{Optical bistability in a two-level atom}

The steady-state solution of the density matrix equations of motion describing a self-interacting optically-driven two-level atom are presented in the main text in terms of the population difference $Z\equiv\rho_{22}-\rho_{11}$ and (slowly-varying) coherence $\tilde{\rho}_{21}$ as
\begin{subequations}
\begin{align}
    \frac{4\gamma}{\Gamma}|\Omega|^2 &= -\frac{Z+1}{Z} \sqpar{ \left(\Delta - \Ree\{G\} Z\right)^2 + \left( \gamma - \Imm\{G\}Z \right)^2 } ,  \label{eq:SS_eq_Z} \\
    \tilde{\rho}_{21} &= \frac{Z \Omega}{(\Delta - \Ree\{G\}Z)+\ii(\gamma-\Imm\{G\}Z)} ,
\end{align}
\end{subequations}
where $\Omega$ is the Rabi frequency, $G$ characterizes the self-interaction, and $\Delta$ is the detuning, while $\Gamma$ and $\gamma$ denote the population decay and dephasing, respectively. Note that for $G\neq 0$, Eq.\ \eqref{eq:SS_eq_Z} is a third-order polynomial in $Z$ that admits up to three real solutions. When such three real distinct solutions can be realized, the two-level atom is in a bistability regime that admits two stable solutions and one unstable solution.

The regions of bistability can be identified by examining the discriminant of the third-order polynomial in Eq.\ \eqref{eq:SS_eq_Z}, which is elucidated by writing it in the form
\begin{equation}
    A x = (x+1) \left( x^2 + Bx + C \right) ,
\end{equation}
where $A=-4\gamma|\Omega|^2/(\Gamma |G|^2)$, $B=-2 \left( \Delta \Ree\{G\} + \gamma \Imm\{G\} \right)/|G|^2$, and $C=(\Delta^2+\gamma^2)/|G|^2$; the discriminant of the polynomial has the form \cite{abramowitz1948handbook}
\begin{equation} \label{eq:discriminant}
    D = -4p^3 - 27q^2 ,
\end{equation}
where $p=B+C-A-(1+B)^2/3$ and $q=C+2(1+B)^3/27-(1+B)(B+C-A)/3$. Now, when $D=0$, Eq.\ \eqref{eq:SS_eq_Z} has multiple real roots, while if $D<0$ and the polynomial has only real coefficients (as in this case),  Eq.\ \eqref{eq:SS_eq_Z} has two non-real complex conjugate roots. Finally, a positive discriminant $D>0$ with real-valued coefficients indicates that Eq.\ \eqref{eq:SS_eq_Z} has three real distinct roots. In the main text we compute the discriminant given in Eq.\ \eqref{eq:discriminant} over a range of parameters to check when there are three steady states, two of which will be stable and one unstable according to linear stability analysis. 

\section{Linear stability analysis}

To assess the stability of the steady state solutions, we consider a small perturbation by writing $Z = Z^{\rm ss} + \delta Z$, $\tilde{\rho}_{21} = \tilde{\rho}_{21}^{\rm ss} + \delta \tilde{\rho}_{21}$, and $\tilde{\rho}_{12} = \tilde{\rho}_{12}^{\rm ss} + \delta \tilde{\rho}_{12}$, so that the linearized optical Bloch equations can be written as
\begin{equation} \label{eq:linear_stability_analysis_Jacobian}
    \frac{d}{dt} \begin{pmatrix} \delta Z \\ \delta \tilde{\rho}_{21} \\ \delta \tilde{\rho}_{12} \end{pmatrix}
    = \begin{pmatrix}
        -\Gamma & - 2 \ii \sqpar{(\Omega+G \tilde{\rho}_{21})^* - G \tilde{\rho}_{12}^{\rm ss}} & 2 \ii \sqpar{(\Omega+G \tilde{\rho}_{21} - G^* \tilde{\rho}_{21}^{\rm ss}} \\
        - \ii \Omega+G \tilde{\rho}_{21}) & \ii \Delta - \gamma - \ii G Z^{\rm ss} & 0 \\
        \ii (\Omega+G \tilde{\rho}_{21})^* & 0 & -\ii \Delta - \gamma + \ii G^* Z^{\rm ss}
    \end{pmatrix}
    \begin{pmatrix} \delta Z \\ \delta \tilde{\rho}_{21} \\ \delta \tilde{\rho}_{12} \end{pmatrix} .
\end{equation}
We identify the above matrix as the Jacobian $J$, with eigenvalues $\lambda$ defined by $\text{det}(J - \lambda \mathbbm{1})=0$ that indicate whether the steady state value under consideration is stable or unstable: If the real part of all eigenvalues for a steady state is negative, the steady state is said to be stable, but if one or more eigenvalues exhibits a positive real part then it is deemed unstable \cite{lugiato2015nonlinear}.

\section{Resonance fluorescence}

The Wiener-Khinchine theorem states that the autocorrelation function and the spectrum associated with a stationary random processes are a Fourier pair \cite{orszag2016quantum}. The spectral density of a quantum field can therefore be written as 
\begin{equation}
    S(\ww) = \frac{1}{2\pi}\int_{-\infty}^\infty d\tau \ee^{\ii\ww\tau}\av{\Eb^-(t+\tau)\cdot\Eb^+(t)}_{\rm ss}
    \propto \Ree\clpar{\int_0^\infty d\tau \ee^{\ii\ww t}\av{\sigma^+(t+\tau)\sigma^-(t)}_{\rm ss}} ,
\end{equation}
where $\av{\dots}_{\rm ss}$ denotes the steady-state, $\Eb^+$($\Eb^-$) is the positive(negative) frequency component of the electric field and $\sigma^-$ ($\sigma^+$) is the atomic lowering (raising) operator \cite{meystre2007elements}. Defining vectors $\vec{\rho}=\ccpar{\rho_{22},\rhot_{21},\rhot_{12}}$ and $\vec{K}=\ii\ccpar{0,\Omega+G\rhot_{21},-\Omega^*-G^*\rhot_{12}}$, the optical Bloch equations for the two-level atom can be expressed as the matrix equation
\begin{equation}
    \frac{d\vec{\rho}}{dt} = \Mm\vec{\rho} + \vec{K} ,
\end{equation}
where
\begin{equation}
    \Mm = \begin{pmatrix} -\Gamma & -\ii\ccpar{\Omega^* + G^*\rhot_{12}} & \ii\ccpar{\Omega + G\rhot_{21}} \\ -2\ii\ccpar{\Omega + G\rhot_{21}} & \ii\Delta-\gamma & 0 \\ 2\ii\ccpar{\Omega^* + G^*\rhot_{12}} & 0 & -\ii\Delta-\gamma \end{pmatrix} . 
\end{equation}
We note that the $\Mm$ and $\vec{K}$ still contain the coherence from the mean dipole moment of the TLA, so the matrix equation is essentially nonlinear. The density matrix elements can be recast into the quantum mechanical expectation value of the spin operators, where the spin operators are $\vec{\sigma} = ( \ket{2}\bra{2}, \ket{1}\bra{2}, \ket{2}\bra{1})^{\text{T}}$, such that $\vec{\rho} = \av{\vec{\sigma}}$. The equation of motion of the mean spin operators are then
\begin{equation} \label{eq:EOM_mean_spin_operator}
    \frac{d}{dt}\av{\vec{\sigma}} = \Mm\av{\vec{\sigma}} + \vec{K} .
\end{equation}
In this notation, the spectrum is
\begin{equation}
    S(\ww) \propto \Ree\clpar{\int_0^\infty d\tau\ee^{-\ii\ww\tau}\av{\sigma_3(\tau)\sigma_2(0)}_{\rm ss}} .
\end{equation}
By decomposing the spin operators as $\vec{\sigma} = \av{\vec{\sigma}} + \delta\vec{\sigma}$, where $\delta\vec{\sigma}$ are fluctuations with a vanishing average $\av{\delta\vec{\sigma}}=0$, the spectrum can be expressed as $S(\ww) = S_{\rm coh}(\ww) + S_{\rm inc}(\ww)$, where
\begin{subequations}
\begin{align}
    S_{\rm coh}(\ww) &= \Ree\clpar{\int_0^\infty d\tau \ee^{-\ii\ww\tau}\av{\sigma_3}_{\rm ss}\av{\sigma_2}_{\rm ss}} = \pi\delta(\ww)\abs{\rhot_{12}}^2 ,  \\
    S_{\rm inc}(\ww) &= \Ree\clpar{\int_0^\infty d\tau\ee^{-\ii\ww\tau}\av{\delta\sigma_3(\tau)\delta\sigma_2(0)}_{\rm ss}}
\end{align}
\end{subequations}
are the coherent and incoherent parts of the spectrum. The equation of motion for the fluctuations $\delta \vec{\sigma}$ can be found from Eq.\ \eqref{eq:EOM_mean_spin_operator} by simply substituting $\av{\vec{\sigma}}=\vec{\sigma} -\delta\vec{\sigma}$. It is important to note that this substitution is not made for the mean spin operator in the matrix $\Mm$ or vector $\vec{K}$, since they originate from the classical dipole moment that contains the mean dipole operator or coherence, and any such fluctuations $\delta \vec{\sigma}$ would go to zero when calculating the quantum mechanical average of the dipole moment. The equation of motion for the fluctuations is then
\begin{equation}
    \frac{d}{dt}\delta\vec{\sigma} = \Mm\delta\vec{\sigma} + \vec{F} ,
\end{equation}
with $\vec{F}$ being a noise operator with a vanishing average \cite{meystre2007elements,mohammadzadeh2019resonance}. Invoking the quantum regression theorem, the correlation function is written as
\begin{equation}
    \frac{d}{d\tau}f_i(\tau) = \sum_j \Mm_{ij} f_j(\tau) ,
\end{equation}
where $f_i(\tau)\equiv\av{\delta\sigma_i(\tau)\delta\sigma_2(0)}$. The incoherent spectrum can then be solved for through a Laplace transformation of the above differential equation to yield \cite{meystre2007elements}
\begin{equation}
    S_{\rm inc}(\ww) = \Ree\clpar{\sum_j\left[(\ii\ww\mathbbm{1}-\Mm)^{-1}\right]_{3j}f_j(0)} ,
\end{equation}
where
\begin{equation}
    \vec{f}(0) = \av{\delta\vec{\sigma}(0)\delta\sigma_2(0)}_{\rm ss} = \av{\vec{\sigma}\sigma_2}_{\rm ss} - \av{\vec{\sigma}}_{\rm ss}\av{\sigma_2}_{\rm ss} = \begin{pmatrix} -\rho_{22}\rhot_{21}  \\ -\rhot_{21}^2  \\ \rho_{22}-\abs{\rhot_{21}}^2 \end{pmatrix}_{\rm ss} .
\end{equation}
Finally, using Cramer's rule, the incoherent spectrum can be found analytically as
\begin{equation}
    S_{\rm inc}(\ww) = \Ree\clpar{\frac{(\ii \omega + \Gamma) (\ii\omega - \ii \Delta + \gamma)f_3(0) - \ii \Omega^* \left[ 2\ii \Omega f_3(0)+2\ii \Omega^* f_2(0) \right] + 2 \ii \Omega^* (\ii \omega - \ii \delta + \gamma) f_1(0)}{(\ii \omega + \Gamma) ( \ii \omega - \ii \Delta + \gamma)( \ii \omega + \ii \Delta \gamma) + 4 |\Omega|^2 (s+\gamma)}} .
\end{equation}

\section{Antibunching}

Antibunching is quantified by the second order correlation function
\begin{equation}
    G^{(2)}(\tau) = \langle E^-(t_0)E^-(t_0+\tau)E^+(t_0+\tau)E^+(t_0)\rangle \propto \langle \sigma_+(t_0)\sigma_+(t_0+\tau)\sigma_-(t_0+\tau)\sigma_-(t_0)\rangle,
\end{equation}
where $t_0$ is assumed to be at the steady state \cite{meystre2007elements,mohammadzadeh2019resonance}, and we neglect any contribution from the graphene sheet itself. Using the relation $\sigma_+(t)\sigma_-(t)=\frac{1}{2}(1+\sigma_z)$, where $\sigma_z$ is a Pauli spin matrix, the correlation function may be written as:
\begin{equation}
    G^{(2)}(\tau) \propto \frac{1}{2} \langle \sigma_+(t_0)\sigma_-(t_0)\rangle + \frac{1}{2}\langle \sigma_+(t_0)\sigma_z(t_0+\tau)\sigma_-(t_0)\rangle =\frac{1}{4} (1+\langle \sigma_z(t_0)\rangle) + \frac{1}{2}\langle \sigma_+(t_0)\sigma_z(t_0+\tau)\sigma_-(t_0)\rangle .
\end{equation}
The average of the products of spin operators above can be evaluated by defining the vector
\begin{equation}
    \vec{C}(t_0,\tau) = \begin{pmatrix}
        \langle \sigma_+(t_0)\sigma_z(t_0+\tau)\sigma_-(t_0)\rangle\\
        \langle \sigma_+(t_0)\sigma_-(t_0+\tau)\sigma_-(t_0)\rangle\\
        \langle \sigma_+(t_0)\sigma_+(t_0+\tau)\sigma_-(t_0)\rangle
    \end{pmatrix},
\end{equation}
and invoking the quantum regression theorem (as in the derivation for the resonance fluorescence spectrum) to write
\begin{equation} \label{eq:SI_diff_eq_for_antibunching}
    \frac{d}{d\tau} \vec{C}(t_0,\tau) = \mathcal{B}(t_0+\tau) \cdot \vec{C} (t_0,\tau) + \langle \sigma_+(t_0)\sigma_-(t_0)\rangle \vec{D} ,
\end{equation}
where $\mathcal{B}$ and $\vec{C}$ defined from the optical Bloch equations
\begin{equation}
    \frac{d}{dt} \begin{pmatrix} \rho_{22}-\rho_{11} \\ \rho_{21} \\ \rho_{12} \end{pmatrix} 
    = \mathcal{B}(t) \cdot \begin{pmatrix}
        \rho_{22}-\rho_{11} \\ \rho_{21} \\ \rho_{12} \end{pmatrix} + \vec{D} ,
\end{equation}
leading to
\begin{equation}
    \mathcal{B} = \begin{pmatrix}
    -\Gamma & - 2 \ii \tilde{\Omega}^* & 2\ii \tilde{\Omega} \\ 
        -\ii\tilde{\Omega} & \ii\Delta - \gamma & 0 \\
        \ii\tilde{\Omega}^* & 0 & -\ii\Delta - \gamma
    \end{pmatrix}
\end{equation}
and $\vec{D}=(-\Gamma,0,0)^{\rm T}$, while $\tilde{\Omega} = \Omega + G\rho_{21}$. The correlation function $G^{(2)}(\tau)$ can now be computed by solving Eq.\ \eqref{eq:SI_diff_eq_for_antibunching} with the initial condition $\vec{C}(t_0,0)=\left(\langle\sigma_z(t_0)\rangle-1,0,0\right)/2$.
    
The normalized $g^{(2)}(\tau)$ can similarly be calculated in the steady state limit
\begin{equation}
    g^{(2)}(\tau) = \lim_{t_0\rightarrow \infty} \frac{\langle E^-(t_0)E^-(t_0+\tau)E^+(t_0+\tau)E^+(t_0)\rangle}{\langle E^-(t_0)E^+(t_0)\rangle\langle E^-(t_0+\tau)E^+(t_0+\tau)\rangle} = \lim_{t_0\rightarrow \infty} \frac{\langle \sigma_+(t_0)\sigma_+(t_0+\tau)\sigma_-(t_0+\tau)\sigma_-(t_0)\rangle}{\langle \sigma_+(t_0)\sigma_-(t_0)\rangle^2} .
\end{equation}

\section{Critical slowing down}

\begin{figure} \label{fig:crit_slowdown}
    \centering
    \includegraphics[width=\textwidth]{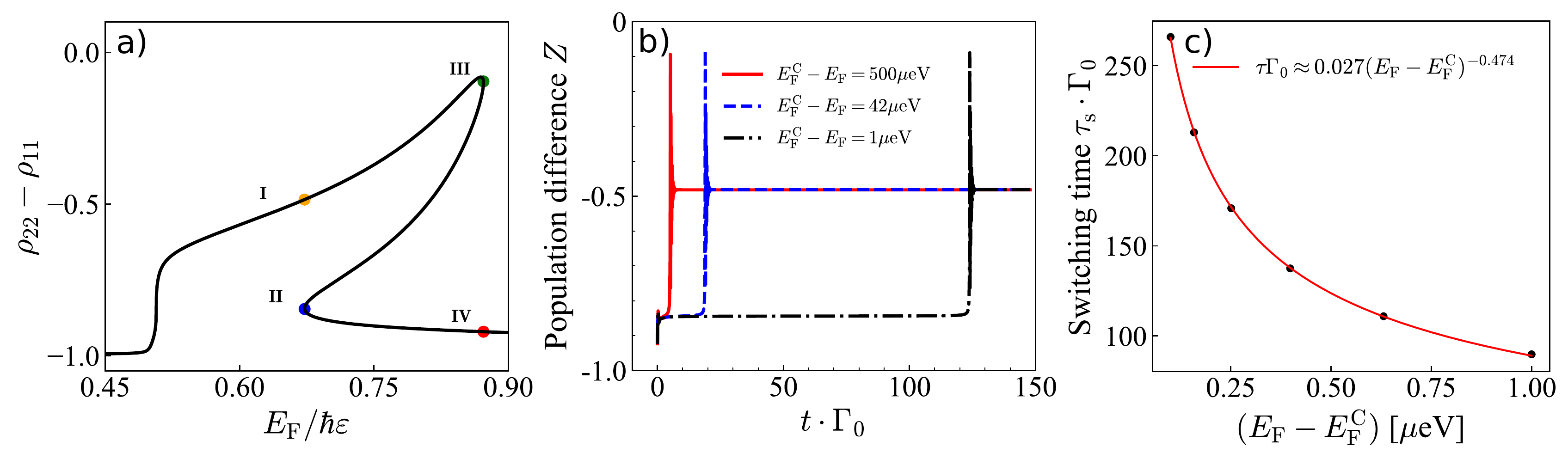}
    \caption{{\bf Critical slowing down.} (a) Steady state population of the two-level atom as a function of $E_{\rm F}/\hbar\vep$ as in Fig.\ 3 of the main paper. (b) Time evolution of the population inversion from $\EF=0.9$\,eV to just below the critical point at point (I) in (a). (c) Switching time from upper stable branch to lower, corresponding from slightly to the left of point (III) to point (IV) indicated in (a).}
\end{figure}

The switching time---the time to go from the lower stable branch to the upper or vice versa---is, for simplicity, in this case set to the time until the maximum value of the population inversion as seen in the maxima in Figure \ref{fig:crit_slowdown}b \cite{nugroho2013bistable}. As noted in the main text, the switching time $\tau$ is not the time until steady state is reached, and small quantitative deviations from the power law are expected as a consequence of the chosen definition of the switching time. In Fig.\ \ref{fig:crit_slowdown}b, one can see that the closer to the critical point the Fermi energy is the longer it takes for the system to reach the steady state. In particular, for the system closest to the critical point, a metastable state can be seen.
The switching time investigated in the text is from a Fermi energy well below the onset of bistability to just above the first bistability region in the phase map. This switching time yielded critical exponents close to $\alpha \approx 0.5$. Similarly, one can go from the upper stable branch to the lower stable branch below. This will yield a similar critical exponent as displayed in Fig. \ref{fig:crit_slowdown}c. The critical exponent is here further from $0.5$, but this may be due to the more imprecise definition of switching time which is here chosen as the minimum population difference.


%